# A study on combined effects of stochastic magnetic fluctuations and synchrotron radiation on the production of runaway electrons

Shucai Li, Lu Wang,* Z. Y. Chen,** D. W. Huang, Weixin Guo, R. H. Tong, F. T. Cui

*State Key Laboratory of Advanced Electromagnetic Engineering and Technology, School of Electrical and Electronic Engineering, Huazhong University of Science and Technology, Wuhan, Hubei 430074, China*

*E-mail: luwang@hust.edu.cn

**E-mail: zychen@hust.edu.cn

## Abstract

The dynamics of relativistic runaway electrons are analyzed using the relativistic Fokker-Planck equation including deceleration due to the synchrotron radiation and radial diffusion loss caused by stochastic magnetic fluctuations (SMFs). SMFs are treated as friction force in [J. Martín-Solís et al., Phys. Plasmas **6,** 3925 (1999)]. However, we think SMFs act as a "porter" in configuration space, but not directly affect the runaway electrons (REs) in momentum space. Both critical electric fields for sustainment of the existing REs and for avalanche onset are enhanced, and the modified avalanche growth rate is reduced by the combined effects of SMFs and synchrotron radiation as compared to the case with only synchrotron radiation [P. Aleynikov et al., Phys. Rev. Lett. **114,** 155001 (2015)].

## 1. Introduction

Nowadays, much attention has been paid to plasma disruption which is almost inevitable in the future tokamak. One of the detrimental effects of disruption is the generation and subsequent loss of energetic runaway electrons (REs). It is well known that in plasmas, the friction force from Coulomb collisions acting on electrons decreases with the increase of velocity, and the electric field can accelerate the electrons. So, runaway will occur when the electrons gain sufficient energy from the electric field to overcome the friction force. The critical electric field for runaway occurring is defined as $E_c = \frac{n_e e^3 \ln \Lambda}{4\pi \varepsilon_0^2 m_e c^2}$ [1], where $n_e$ is the bulk electron density, $e$ is the elementary charge, $m_e$ is the electron rest mass, $c$ is the speed of light in vacuum, $\varepsilon_0$ is the permittivity of free space, and $\ln \Lambda$ is the Coulomb logarithm. The generation mechanism of REs is a hot topic in plasmas research field.

In general, the mechanism of REs generation can be divided roughly into primary generation

[2] (or Dreicer generation), hot tail generation [3] and secondary generation [4, 5] (or avalanche generation). The generation of REs in disruption is a common feature in present day tokamaks [6]. In the process of REs generation, Dreicer generation and hot tail generation usually provide the seed REs, then they are amplified by the secondary avalanche mechanism leading to the exponential growth of runaway population. In addition, runaway seed population can be also generated by tritium decay and Compton scattering of hard x-rays emission from the activated wall [7]. It is widely believed that the secondary generation is responsible for REs in large machines, because it is more effective than the primary generation [8]. For International Thermonuclear Experimental Reactor (ITER), it has been predicted that about two thirds of the pre-disruption plasma current may be replaced by runaway current, mainly due to the avalanche mechanism during the plasma disruption [9]. Since these accelerated relativistic electrons can gain energy up to several tens MeV, they have the potential to produce unacceptable damage to the machine itself if lost from the plasma [10]. However, understanding of runaway generation and runaway loss is still incomplete, and there are inconsistencies between theory and experiment [11, 12].

From theory, the critical electric field $E_c$ for runaway generation is often referred to the critical field for runaway avalanche obtained by the balance between the deceleration by friction force from Coulomb collision and the acceleration by electric field. However, a lot of experiments have observed that the critical electric field is about $(3\sim5)E_c$ [13, 14], and there are many candidates for this disagreement. One of them is synchrotron radiation due to electrons' gyration and guiding-center motion in a curved stochastic magnetic field [15, 16]. For energetic electrons, radiation damping will significantly change their momentum space dynamics. The important influence is runaway electrons cannot be accelerated by the electric field infinitely. In contrast, the runaway energy is limited by synchrotron losses [15]. Due to the exist of energy limit, two different critical electric fields are pointed out [17]. A minimal field $E_0$ is required for sustainment of the existing runaway population, and a higher field $E_a$ is required for the avalanche onset. This leads to a hysteresis behavior in the runaway avalanche. Moreover, the Rosenbluth-Putvinski avalanche growth [5] is reduced because of the energy limit, especially around the critical electric field for avalanche onset.

Another alternative is the anomalous REs loss due to fluctuations of the electromagnetic fields.

It is widely recognized that microturbulent processes play a key role in the anomalous transport in tokamak plasmas. Microturbulence leads to fluctuations of the electric and magnetic fields which may cause an enhancement of REs loss from plasmas. Many theoretical efforts have been dedicated to the studies of the effects of SMFs on REs in tokamak [18, 19]. Ref. [18] describes SMFs as friction force which directly damp electrons in momentum space. Actually, the intact magnetic surfaces may be distorted or destroyed due to the presence of SMFs. Then, there will be a component of magnetic field along the direction of the equilibrium density and temperature gradients. Thus, the radial diffusion loss of REs is strengthened. In this sense, the treatment of SMFs in Ref. [19] is more reasonable, in which the avalanche growth rate of REs is derived by including the radial diffusion loss due to the SMFs. However, the synchrotron radiation is not considered, and the runaway momentum range is from a critical value to infinity in Ref. [19]. As mentioned before, the energy (momentum) limit due to the synchrotron radiation can also affect the avalanche amplification. Therefore, the motivation of our work is to investigate the combined effects of energy limit due to synchrotron radiation and radial diffusion loss by SMFs on REs dynamics.

In the present paper, we consider the combined effects of SMFs and synchrotron radiation, and solve the relativistic Fokker-Planck equation in separated time scales. An approximate analytical formula for the modified avalanche growth rate is derived. Compared with the situation without considering SMFs [17], not only the avalanche growth rate is reduced, but also both two critical electric fields are enhanced. In this work, the sustainment field is determined by the balance between the time scales of momentum evolution and radial diffusion loss induced by SMFs, and is increased with the level of SMFs. These results indicate the important influence of synchrotron radiation and SMFs on the REs in tokamak plasmas.

The remainder of our paper is organized as follows. In the Sec. 2, the relativistic Fokker-Planck equation including both the synchrotron radiation and the SMFs is presented, and the separated time scales are discussed. The modified avalanche growth rate is presented in Sec. 3. Finally, a summary of our work and some discussions are presented in Sec. 4.

## 2. Fokker-Plank equation including both the synchrotron radiation and the SMFs

The REs are governed by the gyrokinetic relativistic Fokker-Planck equation [19, 20],

$$\frac{\partial F}{\partial t} + \frac{\partial}{\partial p}\left[\left(E\xi - 1 - \frac{1}{p^2} - \frac{p\sqrt{(p^2+1)}}{\bar{\tau}_{rad}}(1-\xi^2)\right)F\right] + \frac{\partial}{\partial \xi}\left[E\frac{1-\xi^2}{p}F - \frac{(Z+1)}{2}\frac{\sqrt{p^2+1}}{p^3}(1-\xi^2)\frac{\partial}{\partial \xi}F\right.$$

$$+\frac{1}{\bar{\tau}_{rad}}\frac{\xi(1-\xi^2)}{\sqrt{p^2+1}}F\right] = S + \frac{\tau}{r}\frac{\partial}{\partial r}\left[rD(p)\frac{\partial F}{\partial r}\right]. \tag{1}$$

Here, $F = 2\pi p^2 F_{gc}$, where $F_{gc}$ is the energetic electron guiding center distribution function, $t$ is normalized to relativistic electron collision time $\tau \equiv 4\pi\epsilon_0^2 m_e^2 c^3/(e^4 n_e \ln\Lambda)$ (all the time scales in the following are normalized to $\tau$), $p = \gamma v/c$ is the normalized electron momentum with $\gamma = 1/\sqrt{1-(v/c)^2} = \sqrt{1+p^2}$ being the relativistic factor. $E$ is normalized to $E_c$, and for simplicity, we assume a constant electric field $E$ parallel to the equilibrium magnetic field $B$. This is reasonable on the time scale of REs avalanche. The time evolution of the profile of electric filed over a longer time scale is considered in Ref. [21]. $\xi = p_\parallel/p$ is defined as pitch-angle variable. The subscript $\parallel$ indicate the direction parallel to the magnetic field. The relativistic Fokker-Planck collision operator [1],

$$C(F) = \frac{\partial}{\partial p}\left[\left(1+\frac{1}{p^2}\right)F\right] + \frac{(Z+1)}{2}\frac{\sqrt{p^2+1}}{p^3}\frac{\partial}{\partial \xi}\left[(1-\xi^2)\frac{\partial F}{\partial \xi}\right], \tag{2}$$

including both drag force and pitch-angle scattering was used, where $Z$ is the effective ion charge number. Based on Lorentz-Abraham-Dirac force [22, 23], the radiation reaction force,

$$\langle\frac{dp}{dt}\rangle_{rad} = -\frac{p\sqrt{p^2+1}}{\bar{\tau}_{rad}}(1-\xi^2), \tag{3}$$

$$\langle\frac{d\xi}{dt}\rangle_{rad} = \frac{\xi}{\sqrt{p^2+1}\bar{\tau}_{rad}}(1-\xi^2), \tag{4}$$

was derived [16, 24] and introduced in the continuity kinetic equation [20, 25], where $\bar{\tau}_{rad} = 6\pi\epsilon_0 m_e^3 c^3/(e^4 B^2 \tau)$ is the normalized synchrotron radiation loss time. Here, the radiation force from guiding center motion was neglected, which is justified for that the electron gyro-radius is much smaller the major radius and the electric field is not too large [15]. Note that Eq. (3) is the same as the form in Ref. [15] obtained from Schwinger's synchrotron radiation force [26] for the same conditions. $S$ is the source term of secondary generation of REs by knock–on collision [5]. $D(p)$ is the momentum dependent diffusion coefficient induced by SMFs. In our work, the treatment of SMFs is similar to Ref. [19], but is different from Ref. [18] where SMFs are treated as friction force. In fully stochastic magnetic field, the momentum dependent diffusion coefficient can be written as [27]

$$D = \pi q_0 R_0 v_\parallel \tilde{b}^2/\gamma^5. \tag{5}$$

Here, $q_0$ is the safety factor, $R_0$ is the major radius, $v_\parallel$ is the electron velocity parallel to the equilibrium magnetic field, $\tilde{b} \equiv \tilde{B}_r/B$ is the normalized radial SMFs amplitude. $\pi q_0 R_0 \tilde{b}^2$ represents the field line diffusion coefficient, and the suppression factor $1/\gamma^5$ is attributed to the phase averaging effect of the electron orbits deviating from the flux surfaces. The normalized diffusion time scale is defined as

$$\bar{\tau}_d(p) = \frac{a^2}{j_0^2 D(p)\tau}, \qquad (6)$$

where $j_0$ is the first zero of Bessel function $J_0$, and $a$ is the minor radius. The corresponding normalized diffusion rate is defined as

$$\Gamma_d(p) = \frac{1}{\bar{\tau}_d} = \frac{j_0^2 D(p)\tau}{a^2}. \qquad (7)$$

Solving Eq. (1) directly is very difficult. However, an approximate analytical solution is possible if there is a separation of the time scales. Pitch-angle equilibrium time scale, $\bar{\tau}_p = p^2(1-\xi^2)/(1+Z)$ [16] is much shorter than the acceleration time scale for REs $\bar{\tau}_a = p/(dp/dt)$, which has been discussed in Ref. [17]. The radial diffusion is momentum dependent, and we assume that the diffusion time scale $\bar{\tau}_d$ is on the same order as $\bar{\tau}_a$. This will be discussed later. And, a much longer time scale is the avalanche growth time $\bar{\tau}_{av} \sim \bar{\tau}_a \ln \Lambda$ [19]. Therefore, we have separated time scales: $\bar{\tau}_p \ll \bar{\tau}_a, \bar{\tau}_d \ll \bar{\tau}_{av}$.

Furthermore, for the time scale of pitch-angle equilibrium $\bar{\tau}_p$, the contribution from synchrotron radiation may be negligible for $\bar{\tau}_{rad} \gg 1$ [17]. By balancing the pitch-angle evolution due to the electric field and collision in the Eq. (1), the angular part of distribution function $F$ can be obtained as

$$F = G(p,r,t) \frac{A}{2\sinh A} \exp[A\xi], \qquad (8)$$

with $A(p) \equiv \frac{2E}{Z+1}\frac{p^2}{\sqrt{p^2+1}}$. The governed equation for undetermined function $G(p,r,t)$ can be obtained by integrating Eq. (1) over all pitch angles

$$\frac{\partial G}{\partial t} + \frac{\partial}{\partial p}[U(p)G] = \frac{\tau}{r}\frac{\partial}{\partial r}\left[rD(p)\frac{\partial G}{\partial r}\right]. \qquad (9)$$

Here,

$$U(p) = \frac{dp}{dt} = -\left[\frac{1}{A} - \frac{1}{\tanh A}\right]E - 1 - \frac{1}{p^2} + \frac{Z+1}{E\bar{\tau}_{rad}}\frac{p^2+1}{p}\left[\frac{1}{A} - \frac{1}{\tanh A}\right] \qquad (10)$$

is the acceleration by the electric field against deceleration by both collision and synchrotron radiation which describes the dynamics of energetic electrons in $p$ space [17]. When electric field $E$ is below the critical field $E_0$, $U(p)$ is always negative, which means all the electrons are decelerated, and no REs are generated. When electric field $E$ is above $E_0$, two roots can be determined from $U(p) = 0$. These two roots correspond to the minimum momentum, $p_{min}$ for the electrons overcoming the drag force and the maximum momentum $p_{max}$ due to radiation damping, respectively. We can define a normalized acceleration rate

$$\Gamma_a(p) = \frac{1}{\bar{\tau}_a} = \frac{U(p)}{p}. \tag{11}$$

Fig. 1 shows the normalized acceleration and diffusion rates as a function of momentum for $Z = 5$, $\bar{\tau}_{rad} = 70$, $j_0 = 2.4$, $a = 2.0 \text{m}$, $q_0 = 1$ $R_0 = 6.2 \text{m}$. These parameters are not changed in the following. The red solid line is the normalized diffusion rate for $\tilde{b} = 1.0 \times 10^{-3}$. The discontinuity lines are normalized acceleration rates for different values of the electric field $E$. It is obviously that the normalized acceleration rate increases with the increase of $E$. For well confined electrons, the acceleration rate must be faster than the diffusion rate. Otherwise, they will be lost from the plasma before being accelerated to runaway electron. For the electric field below a certain critical value $E_0'(\tilde{b})$, the acceleration rate is lower than the diffusion rate for all over range of momentum (e.g., blue dot line in Fig. 1), and we can expect that all the runaway electrons will be lost. In contrast, for $E > E_0'$ (purple dot-dashed line), there is a momentum interval satisfied $\Gamma_a > \Gamma_d$. The fast electrons between this interval can be sustained. The role of $E_0'$ is very similar to that of sustainment critical $E_0$ in Ref. [17], but $E_0'$ should be larger than $E_0$ which will be shown later. This is because the required electric field for $\Gamma_a > \Gamma_d > 0$ is higher than that for $U(p) > 0$ (i.e., $\Gamma_a > 0$). Moreover, we can easily determine the runaway momentum region $(p'_{min}, p'_{max})$ from balancing the acceleration rate and diffusion rate, i.e., $\Gamma_a \sim \Gamma_d$. $p'_{min}$ is the minimum momentum for runaway, and $p'_{max}$ is the maximum runaway momentum limited by both the synchrotron radiation and the radial diffusion loss.

The runaway region for different electric field is shown in Figs. 2 (a) and (b). As expected, $p'_{min}$ ($p'_{max}$) decreases (increases) with increasing electric field, i.e., the runaway region becomes wider with the increase of $E$. Note that our runaway region $(p'_{min}, p'_{max})$ is narrower as compared to Ref. [17] without considering SMFs. Particularly, the minimum momentum for runaway is significantly enhanced, but the maximum momentum is less affected. However, the minimum momentum is slightly enhanced as compared to Ref. [18] where both SMFs and synchrotron radiation are included. And, the maximum momentum agrees well with that in Ref. [18].

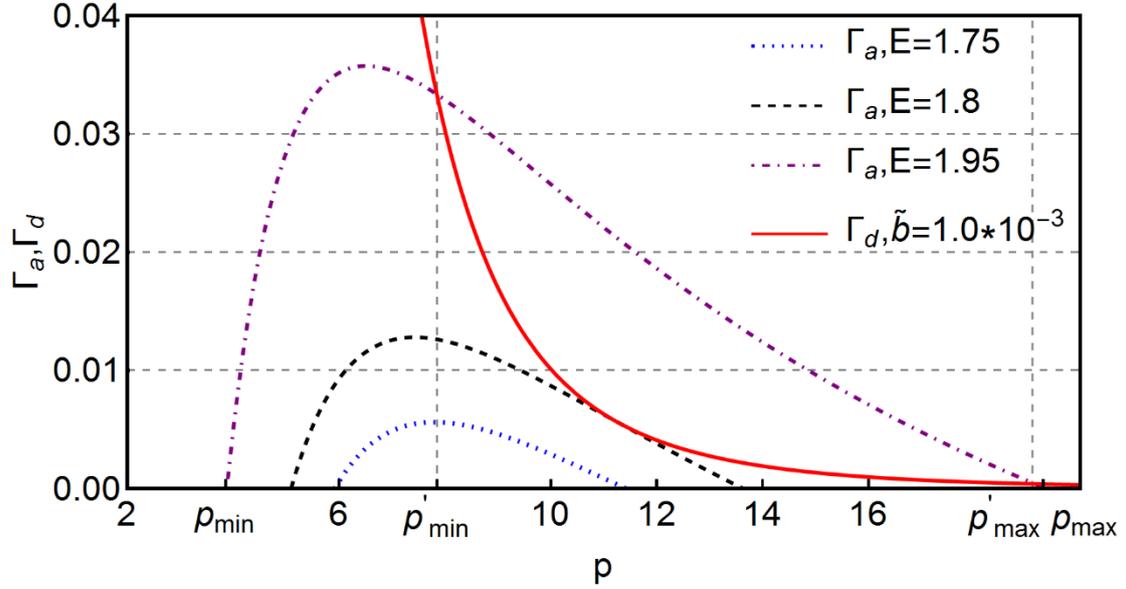

Fig. 1 The normalized acceleration rate and diffusion rate as a function of $p$. The red solid line is the normalized diffusion rate for $\tilde{b} = 1.0 \times 10^{-3}$. The blue dot line, black dash line, and purple dot-dashed line are normalized acceleration rates for the electric field $E = 1.75, 1.8, 1.95$, respectively.

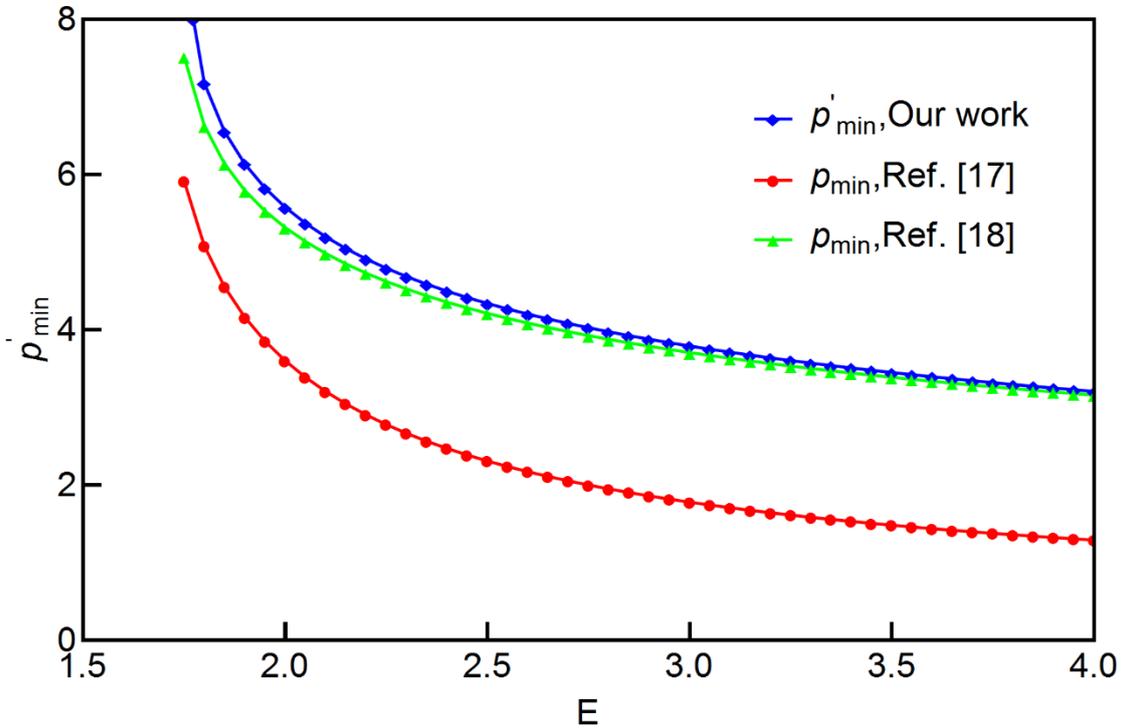

Fig. 2 (a) Minimum runaway momentum $p'_{min}$ (blue diamond line) versus the normalized electric field for $\tilde{b} = 5.0 \times 10^{-4}$. The red circle line and the green triangle line are the corresponding results in Refs. [17] and [18], respectively.

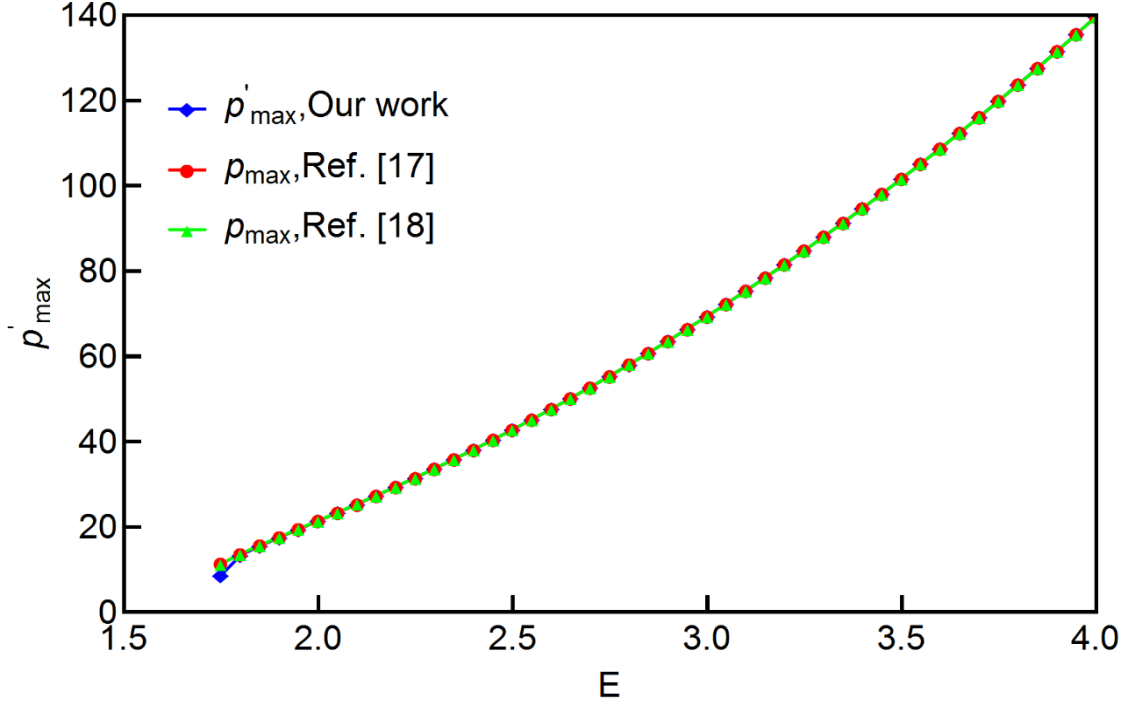

Fig. 2 (b) Maximum runaway momentum $p'_{max}$ (blue diamond line) versus the normalized electric field for $\tilde{b} = 5.0 \times 10^{-4}$. The red circle line and the green triangle line are the corresponding results in Refs. [17] and [18], respectively.

    The effects of SMFs on REs can be understood further through Figs. 3 and 4. Obviously, the runaway region is further narrowed by increasing the level of SMFs. It is found that the SMFs have different effects on the minimum and maximum runaway momenta. For lower energy electron, diffusion coefficient is very sensitive to the SMFs due to smaller orbit deviation from the flux surfaces, and the diffusion loss is very fast. Therefore, the minimum runaway momentum is enhanced a lot as compared to the case without radial diffusion loss [17]. In contrast, electrons with higher energy are less sensitive to the SMFs, and the diffusion loss is very slow. The maximum runaway momentum is thus slightly influenced by the SMFs. Our runaway region agrees with Ref. [18] very well particularly for higher levels of SMFs. Moreover, from Fig. 4, we can see that the critical electric field $E'_0$ required for sustainment of REs is higher than that in Ref. [17]. This is because the acceleration by electric field is required to overcome the radial diffusion loss in addition

to the drag force from collision and synchrotron radiation. The sustainment field $E_0'$ increases with increasing the level of SMFs, which is attributed to the increase of diffusion rate. It can be seen from Fig. 3 and Fig. 4 all the minimum momentum $p_{min}'$, maximum momentum $p_{max}'$, and the sustainment field $E_0'$ of our work will be reduced to the values in Ref. [17] if the SMFs tend to vanish. However, there are small deviations on the minimum momentum and critical electric field between Refs. [17] and [18] for lower levels of SMFs.

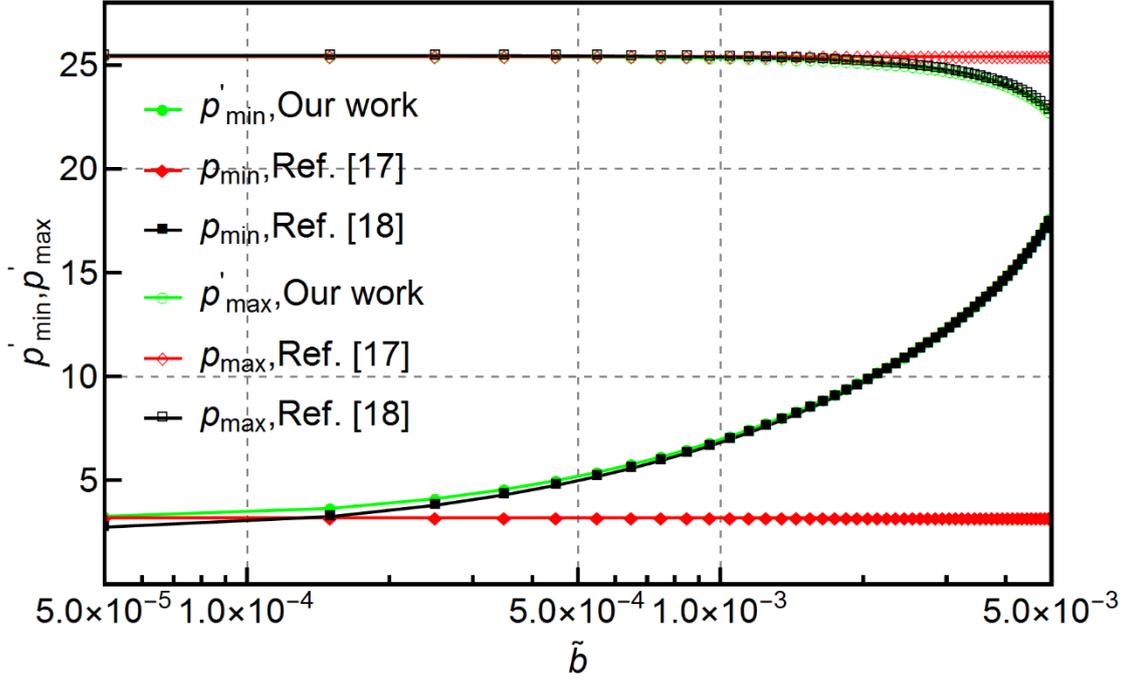

Fig. 3 The minimum and maximum runaway momenta $p_{min}'$ (green filled circle line) and $p_{max}'$ (green empty circle line) as a function of the level of SMFs for $E = 2.1$. The red diamond lines and the black square lines are the corresponding results in Refs. [17] and [18], respectively.

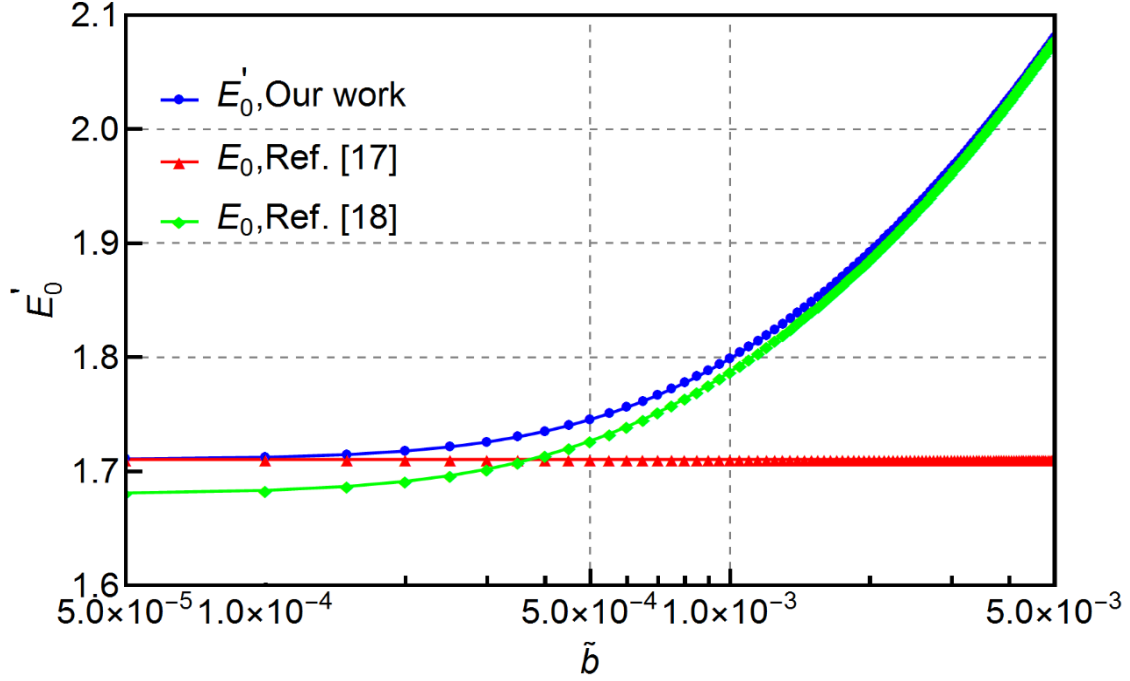

Fig. 4 Critical electric field $E_0'$ (blue circle line) for sustainment of REs as a function of the levels of SMFs. The red triangle line and the green diamond line are the corresponding results in Refs. [17] and [18], respectively.

## 3. Modified avalanche growth rate

Similar to the procedure in Ref. [19], a separable solution of Eq. (9) can be obtained

$$G(p,r,t) = CJ_0(kr)\exp\left[\Gamma t - \int_{p'_{min}}^{p}\frac{\left(\Gamma + \Gamma_d(p') + \frac{dU(p')}{dp'}\right)}{U(p')}dp'\right]$$

$$= CJ_0(kr)\frac{U(p'_{min})}{U(p)}\exp(\Gamma t)\exp\left[-\int_{p'_{min}}^{p}\frac{\Gamma+\Gamma_d(p')}{U(p')}dp'\right]. \quad (12)$$

Here, $C$ is a constant and $k = j_0/a$. $\Gamma$ is the new avalanche growth rate modified by both the synchrotron radiation and the radial diffusion loss. If we take the upper limit of integral as $p'_{min}$, Eq. (12) can be reduced to

$$G(p'_{min},r,t) = CJ_0(kr)\exp(\Gamma t). \quad (13)$$

In addition, the population of the high-energy electrons governed by Eq. (9) is fed from electrons with low momentum by avalanche mechanism. Thus, a boundary condition on $G(p,r,t)$ is written as

$$U(p'_{min})G(p'_{min},r,t) = \Gamma'_r n_r = \Gamma'_r \int_{p'_{min}}^{p'_{max}} G(p)dp, \quad (14)$$

where

$$\Gamma_r' = \frac{1}{4\ln\Lambda\sqrt{\gamma_0^2-1}}\left[-\frac{2\gamma_0-1}{\gamma_0-1}2\ln\left(1+\frac{\gamma_0+1-2\gamma_{mim}}{\gamma_{min}-1}\right)+(\gamma_0+1-2\gamma_{mim})\left(1+\frac{2\gamma_0^2}{(\gamma_{min}-1)(\gamma_0-\gamma_{min})}\right)\right] \quad (15)$$

has the same form as Eq. (11) from Ref. [17]. But, $\gamma_0 = \sqrt{{p'_{max}}^2+1}$ and $\gamma_{min} = \sqrt{{p'_{min}}^2+1}$ are the electron energies corresponding to our new $p'_{max}$ and $p'_{min}$. In Ref. [19], $\Gamma_r'$ is taken as Rosenbluth-Putvinski's growth [5]. Another difference from Ref. [19] is that the runaway momentum region is from $p'_{min}$ to $p'_{max}$, but not from 0 to $\infty$. Here, $U(p'_{max})G(p'_{max},r,t)\approx 0$ was used, since $\Gamma_a(p'_{max}) = \frac{U(p'_{max})}{p'_{max}} \approx 0$ as can be seen from Fig. 1. Combining Eqs. (13) and (14), we can obtain the constant,

$$C = \frac{\Gamma_r' n_r}{U(p'_{min})J_0(kr)\exp(\Gamma t)}. \quad (16)$$

Then, substituting the constant into Eq. (12), we have

$$G(p,r,t) = \frac{\Gamma_r' n_r}{U(p)}\exp\left[-\int_{p'_{min}}^{p}\frac{\Gamma+\Gamma_d(p')}{U(p')}dp'\right]. \quad (17)$$

Integrating Eq. (17) from $p'_{min}$ to $p'_{max}$ and canceling $n_r$, we can obtain

$$1 = \Gamma_r'\int_{p'_{min}}^{p'_{max}}\frac{1}{U(p)}\left\{\exp\left[-\int_{p'_{min}}^{p}\frac{\Gamma}{U(p')}dp'\right]\cdot\exp\left[-\int_{p'_{min}}^{p}\frac{\Gamma_d(p')}{U(p')}dp'\right]\right\}dp. \quad (18)$$

In order to facilitate the integration, the integral upper limit of the third integration on the RHS of Eq. (18) can be extend to the maximum runaway momentum $p'_{max}$. This simplification is reasonable for $\frac{\Gamma_d(p')}{U(p')} = \frac{\Gamma_d(p')}{\Gamma_a(p')p'} \sim \frac{1}{p'}$ which decreases with $p'$ leading to rapid convergence. Thus,

$$\Gamma_r'\int_{p'_{min}}^{p'_{max}}\frac{1}{U(p)}\exp\left[-\int_{p'_{min}}^{p}\frac{\Gamma}{U(p')}dp'\right]\cdot\exp\left[-\int_{p'_{min}}^{p'_{max}}\frac{\Gamma_d(p')}{U(p')}dp'\right]dp$$

$$= \Gamma_r'\exp\left[-\int_{p'_{min}}^{p'_{max}}\frac{\Gamma_d(p')}{U(p')}dp'\right]\int_{p'_{min}}^{p'_{max}}\frac{1}{U(p)}\exp\left[-\int_{p'_{min}}^{p}\frac{\Gamma}{U(p')}dp'\right]dp$$

$$= \frac{\Gamma_r'}{\Gamma}\exp\left[-\int_{p'_{min}}^{p}\frac{\Gamma_d(p')}{U(p')}dp'\right]\left\{1-\exp\left[-\int_{p'_{min}}^{p'_{max}}\frac{\Gamma}{U(p')}dp'\right]\right\}. \quad (19)$$

It follows that Eq. (18) can be written as

$$\Gamma = \Gamma_r'\exp\left[-\int_{p'_{min}}^{p'_{max}}\frac{\Gamma_d(p')}{U(p')}dp'\right]\left\{1-\exp\left[-\int_{p'_{min}}^{p'_{max}}\frac{\Gamma}{U(p')}dp'\right]\right\}. \quad (20)$$

If we make further simplification, the second exponent can be ignored for higher $\Gamma$. This overestimates the growth rate near the threshold. However, the previous extension of integral upper limit underestimates the growth rate. These two approximations may partly compensate for each other. Based on above analysis, an approximate analytical modified avalanche growth rate can be

obtained

$$\Gamma = \Gamma_r' \exp\left[-\int_{p'_{min}}^{p'_{max}} \frac{\Gamma_d(p')}{U(p')} dp'\right]. \tag{21}$$

This can be easily reduced to the case with only synchrotron radiation or only radial diffusion loss considered.

Fig. 5 presents the modified avalanche growth rates as a function of the electric field for different levels of SMFs. The dot-dashed lines and diamond lines are $\Gamma_r'$ and $\Gamma$ determined by Eqs. (15) and (20), respectively. The growth rates are decreased with the increase of levels of SMFs. In order to compare with previous works, the avalanche growth rates in Refs. [5, 17] are also plotted in Fig. 5. Obviously, $\Gamma_r'$ is lower than the predictions by Eq. (18) from Ref. [5] (green dot-dashed line) and Eq. (11) from Ref. [17] (black dash line) which are mainly attributed to the narrower runaway momentum region. The modified avalanche growth rate $\Gamma$ is further weakened by an exponent induced by radial diffusion loss. Moreover, the critical electric field $E_a'$ for the onset of avalanche growth is also enhanced as compared to $E_a$ in Ref. [17], and is increased with increasing the levels of SMFs.

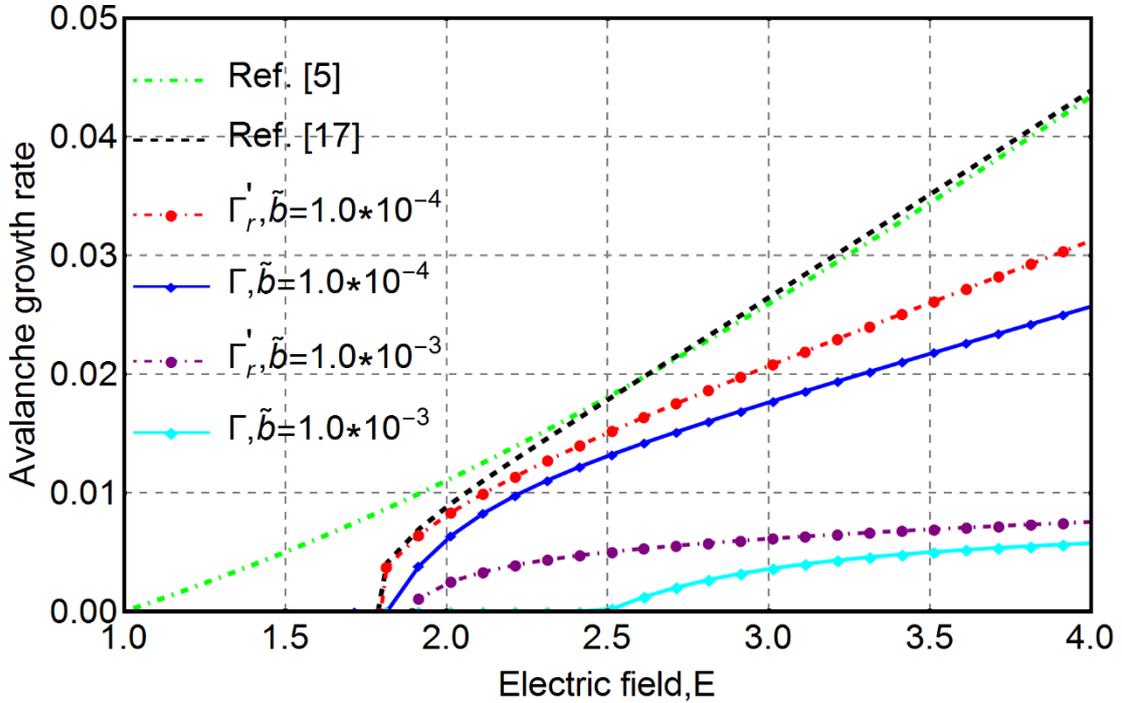

Fig. 5 Modified avalanche growth rates $\Gamma_r'$ (red and purple dot-dashed lines) and $\Gamma$ (blue and cyan diamond lines) predicted by Eqs. (15) and (20), respectively. The green dot-dashed line and black dash line represent Eq. (18) from Ref. [5] and Eq. (11) from Ref. [17], respectively.

## 4. Summary and discussions

The dynamics of relativistic electrons are analyzed using the relativistic Fokker-Planck equation including both the synchrotron radiation and the SMFs. In this work, we treat the effects of SMFs as radial diffusion loss in configuration space. An approximate analytical expression for the modified avalanche growth rate is derived. Both the critical fields for sustainment of REs and for avalanche onset are enhanced, and are increased with the levels of SMFs. The modified avalanche growth rate is much lower than that without considering radial loss by SMFs, and can be significantly lowered by the increase of levels of SMFs. All these results indicate that SMFs and synchrotron radiation have important influences on the REs production and loss, which should be considered for predictions of RE generation and mitigation in future devices such as ITER.

The influence of magnetic fluctuations on the behavior of REs is also confirmed by simulation and experiments. In simulation, the current of runaway beam is decreased with the increase of stochastic magnetic perturbations [28]. In TEXTOR disruptions, the runaway losses were enhanced by the application of resonant magnetic perturbations (RMP) with toroidal mode number n=1 and n=2 [29]. At sufficiently high perturbation levels, a reduction of the runaway current, a shortening of the current and suppression of high energetic runaways were observed. The effects of RMP with an m /n = 2 /1 mode on runaway generation during disruptions were also investigated in J-TEXT [30]. The runaway current was partially suppressed by the application of a magnetic perturbation with $\tilde{b} = 1.3 * 10^{-4}$ during the disruption as compared to the reference discharge without the application of RMP. The influence of intrinsic magnetic turbulence on the loss of REs has also been observed in TEXTOR [31]. For the level of magnetic turbulence larger than $\tilde{b} = 10^{-3}$, the REs were suppressed. Our findings on the enhancement of critical electric field and reduction of avalanche growth rate caused by SMFs are qualitatively in agreement with these simulation results and experimental observations.

**Acknowledgments**

We are grateful to Q.M. Hu, Z.H. Jiang and the participants in the 58th Annual Meeting of the APS Division of Plasma Physics, San Jose, CA, October 31-November 4, 2016 for fruitful discussions. This work was supported by NSFC Grant Nos. 11305071, 11675059, 10275079 and the Ministry of Science and technology of China, under Contract Nos. 2013GB112002,